\documentclass[11pt]{article}

\usepackage[final]{acl}

\usepackage{times}
\usepackage{latexsym}
\usepackage{booktabs}

\usepackage{listings}
\usepackage[T1]{fontenc}

\usepackage[utf8]{inputenc}

\usepackage{microtype}

\usepackage{inconsolata}

\usepackage{graphicx}
\usepackage{soul} 

\title{Relevance is \textit{not} enough: A Communication-Oriented Retrieval System for Consequential Scientific Question Answering}

\author{Avina Nakarmi \\
  New Jersey Institute of Technology \\
  Newark, New Jersey \\
  \texttt{an778@njit.edu} \\\And
  Naga Datha Saikiran Battula \\
  New Jersey Institute of Technology \\
  Newark, New Jersey \\
  \texttt{nb547@njit.edu} \\\AND
  Anthony Diaz \\
  Newark Water Coalition \\
  Newark, New Jersey \\
  \texttt{\small anthony@newarkwatercoalition.org} \\\And
  Aritra Dasgupta \\
  New Jersey Institute of Technology \\
  Newark, New Jersey \\
  \texttt{aritra.dasgupta@njit.edu} \\}

\begin{document}
\maketitle
\begin{abstract}
AI systems increasingly answer scientific questions about health, safety, and the environment. But most retrieval-augmented generation systems are tuned to provide factually correct, on-topic answers rather than to help non-experts understand what those answers mean for their lives and decisions. We focus on consequential scientific questions whose results directly shape people's lives and study them through a public water-quality communication system, where residents and community leaders interpret the findings and choose actions. Their experiences show that on-topic answers can still be insufficient without explanation and context and that the emotional weight of risk information cannot be ignored.
Our system first classifies each question by reasoning type (for example, causal versus policy-based), then generates follow-up questions to identify missing evidence and retrieve it. One component clearly distinguishes between what is known and what is uncertain, while another rewrites scientific details into accessible language, using persona-based styles, such as a caring neighbor or an administrative official, to adapt tone and readability.
Ablations on over $160$ questions show that the system uses \textit{an order of magnitude less context} and, in several configurations, improves human-rated completeness. A completeness metric co-designed with community members and a fine-tuned learned judge reveal that standard relevance scores explain about $1\%$ of variation in human completeness ratings, and even the tuned judge only moderately aligns with humans, indicating that completeness is a distinct human-centered objective that current metrics do not reliably capture.
\end{abstract}

\section{Introduction}

Retrieval-augmented generation (RAG) models have achieved strong results on question answering, with much work improving factual correctness, retrieval accuracy, and evidence support~\cite{es2024ragas,kachuee2025prismrag,tamber2025benchmarking}. Yet in many real-world domains, factual correctness is only one requirement. Users also need answers that help them interpret scientific data, understand uncertainty, and decide what to do. We call this broader challenge consequential scientific question answering and study it through water quality communication, where citizens ask whether a measured lead level is harmful and requires action. A correct answer may report the concentration, but a meaningful one must also explain how it relates to regulatory limits, potential health effects, remaining uncertainties, and available options. This highlights a core problem in current retrieval pipelines: even semantically relevant evidence can be insufficient for answering.

A similar issue appears in evaluation. Automated scoring typically emphasizes faithfulness, answer relevance, and contextual relevance~\cite{es2024ragas}, while large language model judges are increasingly being used as human-like evaluators~\cite{desmond2025evalassist,xu2025does,kim2026augmenting,zhang2026evaluating}. In consequential communication, however, these approaches may fail to detect whether answers are complete, transparent, and meaningful for non-experts. To address this, we design a communication-oriented information retrieval system to support water quality outreach, grounded in authoritative local context, ensuring responses are evidence-based, clear, and appropriate for high-concern situations. Water quality is one instance of a broader class of consequential scientific QA problems, showing that responses must change when the goal is not only correctness but communication that supports public understanding and action.
We make three specific contributions. First, we design a reasoning-guided retrieval mechanism that identifies evidential gaps and seeks sufficient rather than merely relevant material, building on iterative and adaptive retrieval~\cite{guo2025beyond,jiang2023active,asai2024self}. Second, we develop a communication-focused response pipeline that separates evidence-based claim generation from audience-oriented formulation and supports personas such as a caring neighbor and a city hall translator, informed by work on empathy and audience adaptation~\cite{sharma2020computational,august2024know,jiang2025jre}. Third, we propose completeness as a human-centered performance criterion and operationalize it through human annotations and model fine-tuning, drawing on user-centric evaluation frameworks. Our experiments with 160 outreach questions show that the system can generate responses with about an order of magnitude less retrieved text and, in some settings, increasing human-rated completeness. However, relevance-based metrics account for about $1\%$ of the variance in human-assessed completeness, and a fine-tuned automated judge achieves only a moderate correlation with human assessments.
\section{Related Work} \label{sec:related_works}

Most retrieval-augmented generation (RAG) systems share a common backbone~\cite{huang2024survey}: a retriever pulls documents that match the query, and a generator turns that context into an answer. Much of the field has focused on improving retrieval through dense passage retrieval~\cite{karpukhin2020dense} and stronger re-ranking~\cite{huang2024survey}, while surveys note both opportunities and challenges for such architectures in information retrieval settings~\cite{breuer2025large}. Underlying this progress is a quiet assumption that topically relevant documents are enough to produce a good answer. We find that this assumption breaks down in consequential scientific communication, where relevance and sufficiency pull apart. A system can surface documents that are semantically related to a question but operationally beside the point, yielding answers that remain faithful to the source but leave the reader unable to understand the situation or act on it.

\par \noindent \textbf{Communicating consequential science to non-experts.} A growing body of work uses language models to make scientific information accessible to lay audiences, generating plain-language summaries~\cite{august2024know} and adapting science journalism to general readers~\cite{jiang2025jre}. Closer to our setting, conversational agents have been built for consequential domains, including water quality education and communication~\cite{ravindran2025application} and medical question answering~\cite{wang2025medcot}, and prior work stresses that credibility and trust shape how non-experts take up such information~\cite{rabjohn2008examining}. These systems advance accessibility, readability, and factual accuracy, but they treat a correct and readable answer as the endpoint. None of the audits assess whether the retrieved evidence is \textit{sufficient} for the reader to understand the situation and act, which is the gap our prognostic and diagnostic checks target.

\par \noindent \textbf{Iterative and adaptive retrieval.} Recent work has moved beyond the strict retrieve-then-generate approach, making the process more flexible and self-directed. Some systems trigger extra retrieval through reflection signals~\cite{asai2024self} or when model confidence drops~\cite{jiang2023active}. Others break complex questions into sub-questions, each with its own retrieval step~\cite{ammann2025question}. Research on graph-structured retrieval for multi-hop reasoning shows that finding relevant passages is not enough; the system also needs intermediate reasoning to connect evidence~\cite{guo2025beyond}. Another line of work prompts or fine-tunes models to lay out reasoning steps or fill requirement templates before answering~\cite{he2025enhancing,wang2025medcot,wang2025CoRAG}. These methods can yield more thoughtful answers, but they require large annotated datasets, which are often slow, expensive, or impractical to collect in specialized fields like water quality~\cite{palinkas2015purposeful}. Instead of teaching the model to deliberate through supervision, we focus on retrieval: determining what reasoning the question requires, identifying gaps in the evidence, and retrieving missing information so that the context is sufficient, not merely relevant.

\par \noindent \textbf{Evaluation metrics.} Standard RAG metrics compare a response to a fixed reference answer. This works when there is only one correct answer, but it falls short for consequential scientific questions where several answers may be valid, and what matters is whether the answer addresses the questioner's needs. LLM-as-a-judge approaches and related automatic frameworks score faithfulness and relevance without manual labeling~\cite{es2024ragas,desmond2025evalassist,xu2025does,kim2026augmenting,zhang2026evaluating}, but growing evidence shows bias in these judges and limited sensitivity to the practical usefulness of answers for human users. In response, the community has turned to more human-centered approaches, including human-in-the-loop evaluation~\cite{desmond2025evalassist}, evaluations that account for both efficacy and satisfaction~\cite{bauer2025manifesto}, and user-centered guidelines~\cite{muller2025advancing}. Beyond coverage of relevant information, effective communication in high-stakes settings also depends on how uncertainty is conveyed: research on communicating scientific uncertainty~\cite{fischhoff2014communicating} and on its effects on public audiences~\cite{gustafson2020review} shows that non-experts reason and decide differently depending on how clearly uncertainty is expressed. This motivates treating transparent uncertainty not as an optional add-on but as part of what makes an answer complete. We therefore define completeness as a human-centered goal for judging responses in consequential scientific communication, one that jointly captures informational coverage and the transparent communication of uncertainty~\cite{voci2024sustainability}.

\begin{figure*}
    \centering
    \includegraphics[width=.84\linewidth]{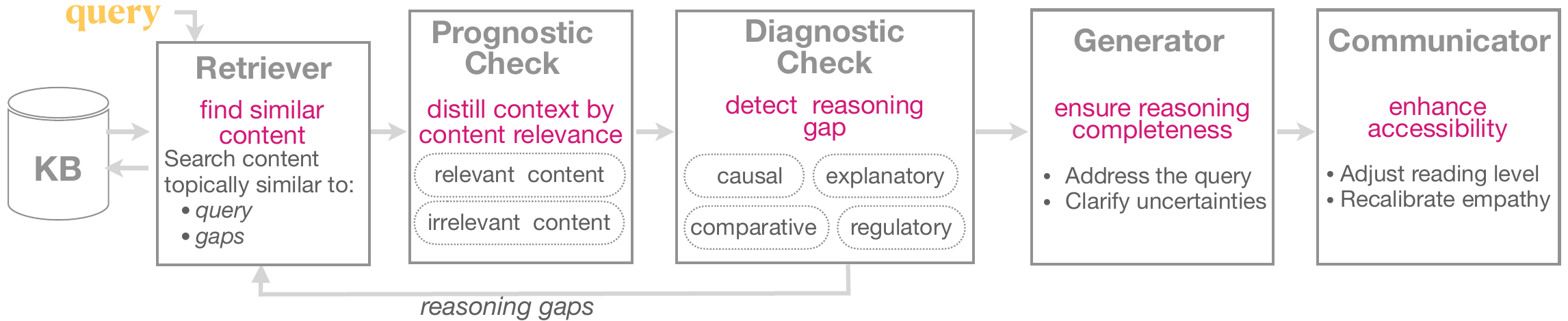}
      \vspace{-0.1in}
    \caption{\textbf{Our community-oriented RAG pipeline} anchors on iterative reasoning gap-driven retrieval from the knowledge base~(KB) for sufficient content extraction and empathy-aware response articulation. Two-step audit using \textit{Prognostic} and \textit{Diagnostic} check controls conciseness and sufficiency of content, respectively.}
    \label{fig:pipeline}
    \vspace{-0.2in}
\end{figure*}



\section{Methodology and System Design}

We designed a RAG-based system for water quality outreach that answers consequential scientific questions requiring contextual explanation, not facts alone. Surveys with affected residents and semi-structured interviews with community leaders disseminating information shaped a communication-oriented pipeline (Figure~\ref{fig:pipeline}) that iteratively builds sufficient context per question, audits retrieved content for precision and sufficiency, and articulates evidence-grounded findings for laypersons.
\subsection{Constructing the knowledge base}
The system operates over a structured knowledge base built from authoritative sources on water quality, contamination regulations, and public health guidance. Scraped documents are stored in a hierarchical JSON structure, with each chunk indexed by its section heading and subheading, so the retriever can match queries against compact, topic-focused headings before falling back to full text.
To probe the pipeline, we assembled an evaluation set of questions. We manually wrote a seed set of frequently asked questions (FAQs) from our community interactions and added questions from the FAQ portals of authoritative public health institutions. We then prompted a large language model (GPT-5.2) to role-play a layperson concerned about tap water, expanding the seed into 500 unique questions used to probe all pipeline variants.
\subsection{Retrieving content}
The retriever embeds chunks with a sentence-transformer model (all-MiniLM-L6-v2) and first matches the user query against the topic-focused headings, which align well with short, user-initiated questions. If no heading exceeds a cosine similarity of 0.51, it falls back to full-text matching, trading precision for broader coverage. Across iterations, the system keeps a running history of retrieved chunks to avoid redundancy and support later gap-filling.
\subsection{Progressively building sufficient context}
Better retrieval does not automatically produce better answers. We observed two failure modes: retrieval includes documents that are semantically similar but unnecessary, inflating context and distracting the generator, and it misses documents whose surface form does not match the query yet are essential to an adequate response, such as regulatory thresholds or background explanations. We address both by splitting retrieval auditing into a prognostic check, which removes low-utility documents to curb contextual overload, and a diagnostic check, which identifies evidential gaps the retrieved content does not cover. Together they balance precision and sufficiency.

\par \noindent \textbf{Prognostic check.} The prognostic check audits retrieved documents before generation. An oracle LLM (gpt-4o-mini) labels each chunk as relevant or irrelevant to the query (Appendix~\ref{app:prog}). We retain both labels, since chunks that appear irrelevant at first may still help close reasoning gaps later. Only relevant chunks advance to the diagnostic check, which reduces contextual noise and the amount of text passed downstream.

\par \noindent \textbf{Diagnostic check.} Relevant documents may still omit critical evidence when topical similarity is low. The diagnostic check evaluates the retrieved content against the evidence required by the query. From community interactions, we identified four recurring reasoning strategies with distinct evidential requirements. \textit{Causal} queries ask why a problem exists and require causes, mechanisms, and confounders. \textit{Regulatory} queries ask which rules or thresholds apply and require relevant policy text, compliance status, and practical implications. \textit{Comparative} queries ask how current conditions differ from past measurements or standards and require reference points and the significance of the difference. \textit{Explanatory} queries ask about unfamiliar terms and require definitions, analogies, and context for non-experts.
We operationalize each strategy as a set of diagnostic questions (Appendix~\ref{app:templates}). For each query, the system classifies its reasoning strategy, checks the retrieved content against the corresponding diagnostic questions, and flags any unanswered questions as evidential gaps. It first tries to fill gaps from previously discarded chunks; for those that remain, it generates targeted sub-queries and runs a second retrieval pass (Appendix~\ref{app:diag}). Because added retrieval rounds offer diminishing returns, we cap this at two iterations. Relevant fetches from the second iteration replace the irrelevant first-iteration chunks, keeping the overall context size stable.

\subsection{Curating responses}
The pipeline grounds every response in the consolidated documents from both retrieval iterations. But factual accuracy alone is insufficient for effective science communication~\cite{rabjohn2008examining}: as our interviews confirmed, responses must also acknowledge the inquirer's sentiment, convey empathy~\cite{voci2024sustainability}, and surface epistemic uncertainty. We therefore split response generation into two stages~\cite{kang2024self}, one focused on evidential grounding and one on accessible, empathetic communication.
\textbf{Generator LLM.} Implemented with GPT-5.2, the Generator LLM takes the relevant documents and identified epistemic gaps and produces a structured set of claims that are logically derived from the evidence and directly address the query. It states epistemic gaps explicitly as knowledge limitations rather than omitting them. This claim-evidence structure keeps responses traceable and surfaces uncertainty, supporting trust in high-stakes communication.
\textbf{Communicator LLM.} The claims, supporting evidence, and stated limitations pass to the Communicator LLM, also GPT-5.2 (Appendix~\ref{app:outputs}). Because LLM-generated text is often hard to read~\cite{roegiest2024generative}, and reading above an audience's level reduces comprehension and engagement~\cite{jiang2025jre}, it rewrites the claims into persona-adapted language while preserving factual content. Our interviews showed that residents often seek information from neighbors or public agencies, so we implement two personas. The \textit{caring neighbor} acknowledges concerns, uses plain conversational language, normalizes uncertainty, and frames actions in manageable terms. The \textit{city hall translator} adopts a civic framing, clarifying which agencies are responsible and how public processes work, and using a neutral tone for uncertainties such as pending decisions or incomplete data.
This modular design, pairing smaller models for retrieval auditing with a stronger shared model for generation and communication, supports deployment by separating lightweight reasoning checks from heavier response synthesis, while keeping responses sufficient, transparent, and emotionally appropriate for real-world use.

\section{Evaluation}
We characterize retrieval performance and response quality using established evaluation metrics for context relevance~\cite{es2024ragas}, answer relevance~\cite{es2024ragas}, emotional reaction~\cite{sharma2020computational}, and reading level~\cite{kincaid1975derivation}. These metrics capture important dimensions of retrieval quality, semantic alignment, emotional framing, and linguistic accessibility.

\par \noindent \textbf{Rationale}. The established metrics do not assess whether a response provides sufficient information for a layperson to understand the issue and its implications. To address this limitation, we co-designed \textit{completeness} as an evaluation metric with the community stakeholders. Moving beyond relevance and accuracy, completeness measures whether a response includes all information necessary for the inquirer to understand the significance, while transparently surfacing epistemic uncertainties.

Uncertainty disclosure is folded into completeness by design rather than scored as a separate dimension. This is a deliberate choice: in a consequential setting, what a reader needs to know is not just the facts but their limits, in other words, the uncertainty associated with events or findings. An answer that presents facts rather than uncertainty is not only less transparent, but also less complete because it withholds information readers need to calibrate how much to trust it and how to act. Silence about what remains unknown leaves readers more confident than the evidence warrants, a failure mode our community collaborators were most concerned about. Treating uncertainty as a separate axis would imply that an answer could be complete in itself while being opaque about its own limits, a decomposition that does not hold for our usage context. We therefore treat transparent communication of uncertainty as a component of completeness rather than as something adjacent to it.

\par \noindent \textbf{Annotation Setup}. Two annotators evaluated 648 pipeline-generated answers across seven variants (a baseline RAG system and six experimental configurations crossing three ablation designs with two target personas) using a five-point Likert scale. Higher scores indicate more complete explanations and transparent uncertainty communication. The annotators achieved substantial agreement (Cohen's $\kappa=0.65$), with disagreements primarily on undefined jargon and poor uncertainty reporting. They penalized overconfidence and a lack of accessible definitions.
These annotations were aggregated into a ground-truth dataset for human evaluation and used to train an LLM judge. {To improve upon low default alignment ($\rho = 0.41$), }we fine-tuned Qwen 2.5-7B on a 60:20:20 (train/val/test) split using Low-Rank Adaptation (LoRA) to minimize computational demand~\cite{hu2022lora}. Training took $\approx35$ minutes over $5$ epochs on a single NVIDIA A100 (80GB) GPU, using approximately $39$ GB of VRAM.

\subsection{Ablation study}
{To isolate the contribution of each component, we evaluated four pipeline configurations on 160 frequent community queries drawn from the 500-question probe set; we report metrics on matched output subsets. Retrieval metrics (Table~\ref{tab:ablation_results}) reflect the $89$ queries answered by all four variants. Completeness is evaluated on the $84$ queries common to the three filtered variants, while baseline completeness is reported on the $n=4$ subset with gold annotations (included for reference given the sample size).}

The first experimental configuration represented a traditional single-pass RAG pipeline (Appendix~\ref{app:baseline}), where all the retrieved chunks were directly passed into the generator (GPT-5.2) to synthesize the final response. The second configuration introduced a variant of the proposed pipeline with an oracle LLM (gpt-4o-mini) for prognostic check, which reduced retrieval noise by filtering out irrelevant content.

The other variants of the pipeline introduced a dynamic, two-iteration retrieval flow that uses an open-source model to perform a diagnostic check. In the third variant of the configuration, a Mistral-7B model identified evidential gaps in the retrieval content without explicit diagnostic criteria. The fourth variant operationalized our complete pipeline, in which the Mistral-7B model was explicitly instructed to classify the query and audit the retrieved evidence according to the diagnostic template for the reasoning strategy. The resulting evidential gaps, in both the third and fourth variants, were converted into targeted retrieval queries. This configuration breakdown helped us examine if the model's implicit reasoning capabilities were sufficient to guide effective iterative retrieval.

An important decision in our design was to use heterogeneous models in the pipeline, including both lightweight, open-source (Mistral-7B) and closed-source (GPT-5.2) models. The diagnostic check does not require the full generative capacity of an LLM; thus, offloading it to a light-weight model is intended to reduce latency and inference cost by assigning lightweight reasoning tasks to a smaller model. The generator and communicator modules, which require stronger language generation capabilities for response formulation and articulation, were implemented using GPT-5.2. This separation of responsibilities led to a modular architecture in which individual components can be independently replaced or fine-tuned as needed. It also allowed examination of each component individually to identify which modules are beneficial and which can be simplified.

\subsection{Results}
This section compares the ablation study variants.
\begin{table}[t]
\small
\setlength{\tabcolsep}{2pt}

\begin{tabular}{l|cc|cc}
\toprule
&
\multicolumn{2}{c|}{Retrieval} &
\multicolumn{2}{c}{Answer Quality} \\
\cmidrule(lr){2-3}
\cmidrule(l){4-5}
Method &
Ctx Len$\downarrow$ &
Ctx Rel$\uparrow$ &
Ans Rel$\uparrow$ &
Completeness$\uparrow$ \\
\midrule
Baseline & 56,032 & 0.068 & \textbf{0.916} & 0.600 \\
Prog & 2,243 & \textbf{0.348} & 0.700 & \textbf{0.751} \\
Diag (imp) & \textbf{2,216} & 0.320 & 0.651 & 0.711 \\
Diag (exp) & 2,226 & 0.336 & 0.688 & 0.749 \\
\bottomrule
\end{tabular}
\vspace{-0.1in}
\caption{
Comparison of pipeline variants. Context length (Ctx Len) is the average number of retrieved characters. Context relevance (Ctx Rel) and answer relevance (Ans Rel) are scores in $[0,1]$ adopted from~\cite{es2024ragas}. Completeness represents human-rated completeness on a $1-5$ Likert scale, normalized to $[0, 1]$ to simplify comparison. Arrows indicate whether higher ($\uparrow$) or lower ($\downarrow$) values are preferred. Best values in each column are shown in bold.
}
\label{tab:ablation_results}
\vspace{-0.2in}
\end{table}

\par \noindent \textbf{Context length and answer quality across retrieval variants.}\label{par:distill_ctx} A key effect of our proposed pipeline is the observed reduction in retrieved context size while maintaining output quality. Context length dropped sharply from baseline ($\mu=56,032$ characters) to all retrieval-filtered variants ($\approx2,200-2,240$ characters). Compared to the baseline, all variants exhibited a significant, 25-fold reduction in context size ($\chi^2=151.02$, all adjusted $p<0.001$).
{Importantly, human-rated completeness increased from 0.60 (baseline) to 0.711–0.751 across filtered variants. Post-hoc tests showed that all variants differ significantly from the baseline in answer relevance ($\chi^2=172.25$, $p<0.001$), with the baseline highest at $0.916$ and the filtered variants at roughly $0.65$--$0.70$.} No significant differences were observed among prognostic and diagnostic variants (all adjusted $p>0.85$). Similarly, completeness remained statistically inseparable across variants. 

\par \noindent \textbf{Relationship between relevance metrics and completeness evaluation.}\label{par:rel_comp} 
{As shown in Table~\ref{tab:ablation_results}, prognostic filtering does not improve answer relevance over the baseline, but increases completeness from $0.60$ to $0.75$.} No significant differences were observed in answer relevance among prognostic and diagnostic variants. Context relevance, unlike answer relevance, improved with filtering ($\chi^2=127.14$, $p<0.001$); prognostic filtering improved performance, whereas diagnostic filtering did not yield additional gain. Human evaluation showed no statistically significant difference across variants ($\chi^2=3.10$, $p=0.38$), despite changes in answer and context relevance. Evidently, the standard relevance metrics did not explain completeness. Analysis showed a weak association between completeness and answer relevance ($\rho=0.05$) and context relevance ($\rho=0.09$). The relevance metrics yield low explanatory power ($R^2=0.014$) in predicting human evaluation of completeness. 

\par \noindent \textbf{{Efficacy of diagnostic check in gap resolution depends on the reasoning strategy.}}\label{par:diag_process}
{While the explicit diagnostic check flags gaps in 72.6\% of queries and resolves them in 73.8\% of those cases, the implicit check flags gaps in only 11.9\% of queries and achieves full closure in just 20.0\%. However, in the explicit diagnostic check, the second retrieval pass introduced new gaps in $4$ of the $61$ re-run queries ($6.6\%$). Additionally, resolution efficacy
is strongly coupled to reasoning strategy. Structured regulatory queries are
highly tractable ($86.4\%$ filled), whereas causal queries remain the primary failure mode
($56.8\%$ filled); comparative queries end the second pass with as many open gaps as they began with, though on only $6$ queries. Crucially, filling evidential gaps does not yield downstream completeness gains,
and the coupling to reasoning strategy vanishes in human ratings
($\chi^2=2.69$, $p=0.44$).}

\begin{table}[t]
\small
\setlength{\tabcolsep}{2pt}
\centering
\centering
\begin{tabular*}{\columnwidth}{@{\extracolsep{\fill}}lcccc@{}}
\toprule
& \multicolumn{2}{c}{Queries with gaps} & \multicolumn{2}{c}{Evidential gaps} \\
\cmidrule(lr){2-3} \cmidrule(lr){4-5}
& itr\,1 & itr\,2 & itr\,1 & itr\,2 \\
\midrule
\multicolumn{5}{l}{\textit{Configuration (n)}} \\
Diag (imp) \small(84)  & 10 &  8 & 10 &  8 \\
Diag (exp) \small(84)  & 61 & 16 & 87 & 26 \\
\midrule
\multicolumn{5}{l}{\textit{Diag (exp), by reasoning strategy (n)}} \\
Regulatory \small(13)  & 12 &  2 & 22 &  3 \\
Explanatory \small(30) & 21 &  5 & 26 &  5 \\
Causal \small(35)      & 26 &  8 & 37 & 16 \\
Comparative \small(6)  &  2 &  1 &  2 &  2 \\
\bottomrule
\end{tabular*}
\vspace{-0.1in}
\caption{
\textbf{Evidential gap resolution across reasoning strategies in diagnostic check.} A query runs a second retrieval pass (itr\,2) when
iteration~1 leaves an evidential gap open; gap counts are summed over those
queries. Prognostic filtering performs no gap analysis and is omitted.
In $4$ of the explicit variant's $61$ re-run queries, the second iteration ends with
more open gaps than iteration~1 identified.
}
\label{tab:diag_metrics}
\vspace{-0.2in}
\end{table}

\par \noindent \textbf{Agreement between LLM-based and human evaluations.}\label{par:comp_IRR} We observed a strong misalignment between human and LLM-based evaluations. Despite explicit evaluation instructions, the LLM judge correlated weakly with humans ($\rho=0.41$). In contrast, a fine-tuned LLM judge showed improved alignment with human evaluations ($\rho=0.73$). Regression analysis also showed that answer relevance ($\beta=0.37$, $p=0.14$) did not have a statistically significant effect on its evaluation. Despite this improvement, the fine-tuned model only showed moderate agreement (Cohen's $\kappa=0.54$) with human evaluation. 

\par \noindent \textbf{Effects of persona assignment on articulation.}\label{par:persona} We observed that adding a communicator LLM altered the answer style. Defining personas had a strong effect on emotional expression ($\chi^2=351.97$, $p<0.001$). Answers generated using the \textit{caring neighbor} persona consistently showed empathy (over $50\%$ showing emotional reaction), which was significantly higher than the \textit{city hall translator} persona (non-detectable) and the baseline. Reading scores differed significantly ($\chi^2=238.85$, $p<0.001$), with answers generated by the \textit{caring neighbor} persona achieving substantially higher readability scores ($\mu=51-52$) than the corresponding \textit{city hall translator} persona ($\mu=41-42$) and no persona assigned ($\mu=36$). However, measures of completeness, answer relevance, and context relevance showed no significant differences across the different persona conditions. 

\begin{figure}
    \centering
    \includegraphics[width=1\linewidth]{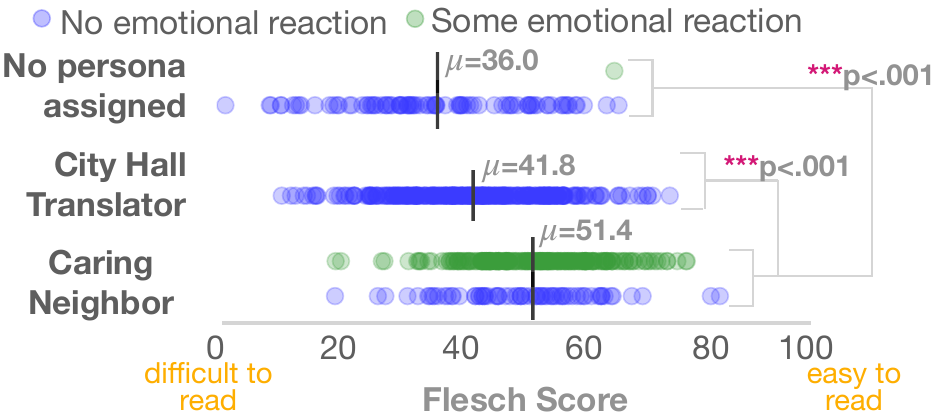}
   \vspace{-0.2in}
    \caption{\textbf{Impact of role-based personas on response readability and empathy.} Each dot marks a response. The caring neighbor persona improved both readability and emotional resonance relative to the no-persona baseline. The "city hall translator" persona also improved readability ($\mu=41.8$ vs.\ $\mu=36.0$) but produced no detectable emotional reaction.}
    \label{fig:Readability_vs_ER}
    \vspace{-0.2in}
\end{figure}

\section{Discussion}
In this section, we analyze the key implications for system design and deployment.

\par \noindent \textbf{Reasoning-guided context compression improves efficiency while preserving quality.} 
As demonstrated in \ref{par:distill_ctx}, filtering achieved a substantial reduction in context length without compromising answer quality. In a practical science communication system, such compression can translate into lower inference cost and reduced latency. Furthermore, minimizing context usage in each interaction allows the system to support longer conversations before reaching context capacity limits.
{Conversely, the explicit diagnostic check yields no statistically significant gain in final completeness over the prognostic-only variant (Table~\ref{tab:ablation_results}), even though it demonstrably detects and closes evidential gaps (\ref{par:diag_process}). Our evaluation set was sampled using message-selection criteria unrelated to reasoning complexity.} In practice, our dataset primarily comprised direct queries with explicit information needs. Future research should explore the utility of diagnostic prompting in ambiguous, open-ended, and multi-hop scenarios.

\par \noindent \textbf{LLM-as-a-judge requires fine-tuning to approximate human evaluation.} 
The initial misalignment between the LLM judge and human annotators (\ref{par:comp_IRR}) underscores that LLM-as-a-judge approaches do not inherently reflect human assessment. However, the improved alignment following fine-tuning suggests automated judges can successfully screen large-scale systems during early development, reducing human annotation workload while reserving final human validation of user-centered qualities before deployment.

\par \noindent \textbf{Answer relevance does not reflect human perception of completeness.} 
The weak association and low explanatory power between relevance metrics and human-rated completeness, as observed in \ref{par:rel_comp}, indicate that responses judged as highly relevant can still omit information humans deem necessary. This mismatch highlights the need to evaluate completeness independently of relevance. In science communication contexts, where users require a holistic view for decision-making, a complete response must complement factual accuracy with the necessary breadth of information.

\par \noindent \textbf{Persona assignment tailors articulation without degrading content.} 
Our findings show that assigning personas significantly alters emotional expression and readability without shifting underlying completeness or relevance scores (\ref{par:persona}). This indicates that persona definitions successfully drive stylistic adaptation while preserving core informational content. Using automated personas allows systems to dynamically tailor information to diverse users, mitigating the risk of rigid, unempathetic responses without requiring labor-intensive manual editing.

\section{Conclusion}
We present a community-oriented RAG system that combines reasoning-guided iterative retrieval with persona-based articulation. The goal of this system is to generate responses that help laypeople interpret consequential scientific information. We introduce \textit{completeness} as a human-centered evaluation criterion and show that it captures aspects of response quality that standard relevance-based metrics overlook. Our results further demonstrate that gap-aware retrieval can substantially reduce context usage. The paper highlights the importance of aligning retrieval and evaluation to human-perceived criteria rather than purely factual objectives in high-stakes scenarios. 

\section*{Limitations}
While our proposed communication-oriented retrieval system improves completeness and reduces context requirements for consequential scientific question answering, several limitations remain. Our pipeline relies on proprietary large language models as black-box components for response generation and communication. As a result, we cannot fully characterize the internal reasoning processes that lead to particular outputs. Furthermore, alternative models or domain-specific fine-tuning could produce different outcomes.

{Moreover, the components of the system transfer across domains unevenly. The core components, including filtering relevant content, identifying gaps, iterative retrieval, and independent articulator, transfer. The four reasoning strategies generalize broadly, allowing domain adaptation primarily through knowledge base substitution. However, the diagnostic questions tailored to water-quality evidence do not transfer and must be customized based on the domain. The water-quality setting should therefore be read as one example application, a case study validating a reusable design rather than a domain-specific system.}

\section{Acknowledgment}
We would like to thank Prof. Neil Maher~(NJIT) and members of the Newark Water Coalition for their help with all stages of this project. This work was supported in part by NJIT's Collaborative Research, Innovation and Strategic Partnerships~(CRISP) grant and the National Science Foundation~(NSF) grant 2312932.

\section*{Ethical Considerations}
Our system is built to assist with communicating critical scientific data, not to replace public agencies or domain experts. Given the highly consequential nature of the information, inaccurate or incomplete information could lead to inappropriate actions. To minimize this risk, the system bases its answers on authoritative sources, clearly highlights areas of uncertainty, and specifies what the evidence does and doesn't prove. However, it utilizes large language models that can hallucinate, omit data, or reflect training biases. While retrieval and auditing components mitigate these risks, they cannot eliminate them, making human oversight essential in high-stakes deployments.

Additionally, the system heavily relies on its knowledge base. While this experiment used strictly curated water-quality and public-health resources, real-world deployments risk institutional bias if not properly supervised. To prevent this, practical applications in consequential domains require transparent governance and strict source-auditing procedures.

Finally, our human evaluation data were collected from annotators assessing response completeness and communication quality. No personally identifiable information was collected or used for model training. The resulting annotations were used solely for research purposes and for developing an automated completeness evaluator.

\bibliography{main}

\appendix
\section{Prompts}
\label{app:prompts}

\subsection{Single-pass RAG baseline (Configuration 1)}
\label{app:baseline}
Generator: \texttt{gpt-5.2}.
\begin{lstlisting}
[system]
Constrain your knowledge to the context given and answer the query

[user]
QUERY: {query}
CONTEXT: {context}
\end{lstlisting}

\subsection{Prognostic check}
\label{app:prog}
Generator: \texttt{gpt-4o-mini}, \texttt{temperature=0}. 
Chunks scoring $\geq 2$ are passed downstream; the rest are retained for later gap filling.
\begin{lstlisting}
[system]
You are a strict evaluator. Score how relevant the given chunk is to the user's query on a 0-3 scale. Return only valid JSON with keys relevance_score.

[user]
    USER QUERY:
    {query}

    RETRIEVED CHUNK:
    Heading:
    {heading}
    Text:
    {text}

    TASK:
    Evaluate whether this chunk helps answer the query.

    Answer in JSON format:
    {
        "relevance_score": 0-3
    }

    SCORING GUIDE:
    0 = irrelevant
    1 = tangentially related
    2 = partially useful
    3 = directly answers query
\end{lstlisting}

\subsection{Diagnostic check}
\label{app:diag}
Generator: \texttt{Mistral:7b}.
\begin{lstlisting}
[system]
You are a judge responsible for analyzing evidence that support specific reasoning. Do not retrieve anything. Do not answer the user's query. Do not evaluate general sufficiency. Only analyze how well the provided evidence satisfies the required diagnostic schema for the given reasoning strategy. Return only valid JSON.

Strategy template:
{reasoning_strategy, diagnostic_questions, failure_modes}
\end{lstlisting}

\noindent Diag (imp) always supplies the \texttt{default} template. Diag (exp) first classifies the query and supplies the matching template.
\begin{lstlisting}
[system]
You are an analyst that identifies the reasoning strategy needed to answer a query. Choose exactly one strategy from: causal, regulatory, comparative, explanatory. Return only valid JSON.
\end{lstlisting}

\subsubsection{Strategy templates}
\label{app:templates}

Default
\begin{lstlisting}
  Q: Are all the relevant concepts, mechanisms, comparisons, or regulations needed to answer the query present and explained clearly in the context of the person's water quality issue?
  F: missing key concept explanation
\end{lstlisting}

Causal
\begin{lstlisting}
  Q: Cause identified — the source or agent must be named.
  Q: Effect described — the observable problem the person is experiencing must be explicitly linked.
  Q: Causal pathway — how the cause produces the effect must be explained in plain terms (not just asserted)
  Q: Temporal precedence — evidence that the cause existed before the problem appeared must be present.
  Q: Confounding factors — other possible causes of the same symptom must be acknowledged.
  Q: Counterfactual or control — evidence from an unaffected comparable source or period must be present.
  F: missing mechanism; missing confounder control; missing temporal evidence; missing enabling conditions; correlation presented as causation
\end{lstlisting}

Regulatory
\begin{lstlisting}
  Q: Concept or policy identified — the regulation, standard, guideline, or regulatory term must be named and placed in its regulatory context.
  Q: Purpose stated — the problem or risk the regulation or standard was designed to address must be present.
  Q: Mechanics explained — how the regulation or standard works in practice must be described (e.g. how MCL is set, how LCR testing is conducted).
  Q: Compliance requirements — what water providers, authorities, or users must do to comply must be stated.
  Q: Changes or evolution — any revisions, updates, or known limitations since introduction must be noted or confirmed absent.
  Q: Implications for the person — what this regulation or standard means for a regular person's safety, rights, or obligations must be made explicit.
  F: missing measured or threshold values; missing policy regulating it; missing synthesis condition; missing connection for isolated evidence
\end{lstlisting}

Comparative
\begin{lstlisting}
  Q: Both comparators named — the two things being compared (sites, periods, sources) must be explicitly identified.
  Q: Shared metrics used — the same parameter measured in compatible units must be present for both comparators.
  Q: Reference or baseline present — a standard, historical baseline, or control site must anchor the comparison.
  Q: Directionality explicit — which comparator is higher, lower, better, or worse must be clearly stated.
  Q: Structural differences disclosed — any meaningful differences between comparators that affect fairness of comparison (e.g. one site is directly downstream of a discharge point) must be acknowledged.
  Q: Reason for the difference — an explanation of why the difference exists, not just that it does, must be present.
  F: missing baseline; missing intervention details; missing comparative metric; alternative explanation ignored
\end{lstlisting}

Explanatory
\begin{lstlisting}
  Q: Clear definition — the jargon, acronym, or concept must be explained in plain, accessible language.
  Q: Scope and application — who or what this definition or rule applies to must be clear.
  F: circular definition; missing baseline or historical context; overly academic or jargon-heavy text; lack of practical relevance; outdated or unverified values
\end{lstlisting}

\section{Representative Outputs}
\label{app:outputs}

All three configurations on one causal query.

\begin{quote}
\small
\textbf{[query]} What is causing lead in drinking water?
\end{quote}

\noindent\textbf{Baseline.}
\begin{quote}
\small
Lead gets into drinking water mainly when plumbing materials that contain lead corrode(wear away/dissolve into the water), especially if the water is more corrosive (for example, higher acidity or low mineral content).
The most common sources mentioned are:
\begin{itemize}
    \item Lead pipes (the most common source), including lead service lines that connect a home to the water main
    \item Brass fixtures (and brass/chrome-plated brass faucets)
    \item Lead solder in older plumbing
    \item For some private wells: “packer” elements in wells 20+ years old, and leaded brass parts in older submersible pumps
\end{itemize}
\end{quote}

\noindent\textbf{Diag (exp), city hall translator.}
\begin{quote}
\small
Lead in drinking water is mainly caused by lead-containing plumbing materials that corrode or leach into the water. The most common contributors are lead pipes, certain older fixtures, and some well components.

In particular, lead pipes are the most common source. Other possible sources include brass fixtures, “packer” elements in private wells that are 20 years old or older, and leaded brass parts in older submersible pumps used in private wells.

The exact amount contributed by each source can vary by home, plumbing system, and well configuration.
\end{quote}

\noindent\textbf{Diag (exp), caring neighbor.}
\begin{quote}
\small
That’s a completely reasonable thing to want to understand — and we can keep it simple.

Lead in drinking water is mainly caused by lead-containing plumbing materials that corrode or leak into the water. The most common sources are lead pipes and certain older fixtures or well parts, including brass fixtures, “packer” elements in private wells that are 20 years old or older, and leaded brass parts in older submersible pumps used in private wells.

What we don’t know exactly is how much each source contributes, because that can vary from home to home, plumbing system to plumbing system, and well setup to well setup.
\end{quote}

\begin{figure}
    \centering
    \includegraphics[width=1\linewidth]{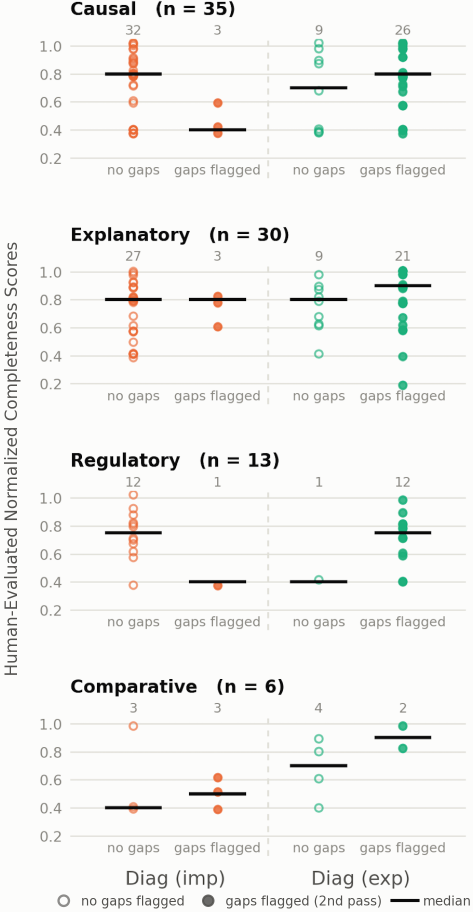}
    \caption{\textbf{Effect of evidential gap filling is statistically insignificant on completeness.} The solid and open marks represent two disjoint query sets, where solid marks indicate queries where flagged gaps triggered a second iteration. The prognostic-only variant is omitted due to its lack of gap analysis and strategy conditioning. The diagnostic variants differ in execution: Diag (exp) explicitly classifies and audits queries against strategy templates, whereas Diag (imp) receives no explicit criteria.}
    \label{fig:placeholder}
\end{figure}

\end{document}